\documentstyle[twoside]{article}

\catcode`\@=11
\long\def\@makefntext#1{
\protect\noindent \hbox to 3.2pt {\hskip-.9pt
$^{{\eightrm\@thefnmark}}$\hfil}#1\hfill}               

\def\@makefnmark{\hbox to 0pt{$^{\@thefnmark}$\hss}}    

\def\ps@myheadings{\let\@mkboth\@gobbletwo
\def\@oddhead{\hbox{}
\rightmark\hfil\eightrm\thepage}
\def\@oddfoot{}\def\@evenhead{\eightrm\thepage\hfil
\leftmark\hbox{}}\def\@evenfoot{}
\def\sectionmark##1{}\def\subsectionmark##1{}}

\oddsidemargin=\evensidemargin
\newcounter{sectionc}\newcounter{subsectionc}\newcounter{subsubsectionc}
\renewcommand{\section}[1] {\vspace{12pt}\addtocounter{sectionc}{1}
\setcounter{subsectionc}{0}\setcounter{subsubsectionc}{0}\noindent
        {\tenbf\thesectionc. #1}\par\vspace{5pt}}
\renewcommand{\subsection}[1] {\vspace{12pt}\addtocounter{subsectionc}{1}
        \setcounter{subsubsectionc}{0}\noindent
        {\bf\thesectionc.\thesubsectionc. {\kern1pt \bfit #1}}\par\vspace{5pt}}
\renewcommand{\subsubsection}[1] {\vspace{12pt}\addtocounter{subsubsectionc}{1}
        \noindent{\tenrm\thesectionc.\thesubsectionc.\thesubsubsectionc.
        {\kern1pt \tenit #1}}\par\vspace{5pt}}
\newcommand{\nonumsection}[1] {\vspace{12pt}\noindent{\tenbf #1}
        \par\vspace{5pt}}

\newcounter{appendixc}
\newcounter{subappendixc}[appendixc]
\newcounter{subsubappendixc}[subappendixc]
\renewcommand{\thesubappendixc}{\Alph{appendixc}.\arabic{subappendixc}}
\renewcommand{\thesubsubappendixc}
        {\Alph{appendixc}.\arabic{subappendixc}.\arabic{subsubappendixc}}

\renewcommand{\appendix}[1] {\vspace{12pt}
        \refstepcounter{appendixc}
        \setcounter{figure}{0}
        \setcounter{table}{0}
        \setcounter{lemma}{0}
        \setcounter{theorem}{0}
        \setcounter{corollary}{0}
        \setcounter{definition}{0}
        \setcounter{equation}{0}
        \renewcommand{\thefigure}{\Alph{appendixc}.\arabic{figure}}
        \renewcommand{\thetable}{\Alph{appendixc}.\arabic{table}}
        \renewcommand{\theappendixc}{\Alph{appendixc}}
        \renewcommand{\thelemma}{\Alph{appendixc}.\arabic{lemma}}
        \renewcommand{\thetheorem}{\Alph{appendixc}.\arabic{theorem}}
        \renewcommand{\thedefinition}{\Alph{appendixc}.\arabic{definition}}
        \renewcommand{\thecorollary}{\Alph{appendixc}.\arabic{corollary}}
        \renewcommand{\theequation}{\Alph{appendixc}.\arabic{equation}}
        \noindent{\tenbf Appendix \theappendixc #1}\par\vspace{5pt}}
\newcommand{\subappendix}[1] {\vspace{12pt}
        \refstepcounter{subappendixc}
        \noindent{\bf Appendix \thesubappendixc. {\kern1pt \bfit #1}}
        \par\vspace{5pt}}
\newcommand{\subsubappendix}[1] {\vspace{12pt}
        \refstepcounter{subsubappendixc}
        \noindent{\rm Appendix \thesubsubappendixc. {\kern1pt \tenit #1}}
        \par\vspace{5pt}}

\newcommand{\textlineskip}{\baselineskip=13pt}
\newcommand{\smalllineskip}{\baselineskip=10pt}

\def\eightcirc{
\begin{picture}(0,0)
\put(4.4,1.8){\circle{6.5}}
\end{picture}}
\def\eightcopyright{\eightcirc\kern2.7pt\hbox{\eightrm c}}

\def\abstracts#1#2#3{{
        \centering{\begin{minipage}{4.5in}\baselineskip=10pt\footnotesize
        \parindent=0pt #1\par
        \parindent=15pt #2\par
        \parindent=15pt #3
        \end{minipage}}\par}}

\newcommand{\bibit}{\nineit}
\newcommand{\bibbf}{\ninebf}
\renewenvironment{thebibliography}[1]
        {\frenchspacing
         \ninerm\baselineskip=11pt
         \begin{list}{\arabic{enumi}.}
        {\usecounter{enumi}\setlength{\parsep}{0pt}
         \setlength{\leftmargin 12.7pt}{\rightmargin 0pt} 
         \setlength{\itemsep}{0pt} \settowidth
        {\labelwidth}{#1.}\sloppy}}{\end{list}}

\newcounter{itemlistc}
\newcounter{romanlistc}
\newcounter{alphlistc}
\newcounter{arabiclistc}

\newcommand{\fcaption}[1]{
        \refstepcounter{figure}
        \setbox\@tempboxa = \hbox{\footnotesize Fig.~\thefigure. #1}
        \ifdim \wd\@tempboxa > 5in
           {\begin{center}
        \parbox{5in}{\footnotesize\smalllineskip Fig.~\thefigure. #1}
            \end{center}}
        \else
             {\begin{center}
             {\footnotesize Fig.~\thefigure. #1}
              \end{center}}
        \fi}

\newcommand{\tcaption}[1]{
        \refstepcounter{table}
        \setbox\@tempboxa = \hbox{\footnotesize Table~\thetable. #1}
        \ifdim \wd\@tempboxa > 5in
           {\begin{center}
        \parbox{5in}{\footnotesize\smalllineskip Table~\thetable. #1}
            \end{center}}
        \else
             {\begin{center}
             {\footnotesize Table~\thetable. #1}
              \end{center}}
        \fi}

\def\@citex[#1]#2{\if@filesw\immediate\write\@auxout
        {\string\citation{#2}}\fi
\def\@citea{}\@cite{\@for\@citeb:=#2\do
        {\@citea\def\@citea{,}\@ifundefined
        {b@\@citeb}{{\bf ?}\@warning
        {Citation `\@citeb' on page \thepage \space undefined}}
        {\csname b@\@citeb\endcsname}}}{#1}}

\newif\if@cghi
\def\cite{\@cghitrue\@ifnextchar [{\@tempswatrue
        \@citex}{\@tempswafalse\@citex[]}}
\def\citelow{\@cghifalse\@ifnextchar [{\@tempswatrue
        \@citex}{\@tempswafalse\@citex[]}}
\def\@cite#1#2{{$\null^{#1}$\if@tempswa\typeout
        {IJCGA warning: optional citation argument
        ignored: `#2'} \fi}}

\def\pmb#1{\setbox0=\hbox{#1}
        \kern-.025em\copy0\kern-\wd0
        \kern.05em\copy0\kern-\wd0
        \kern-.025em\raise.0433em\box0}

\def\fnt#1#2{\footnotetext{\kern-.3em
        {$^{\mbox{\scriptsize #1}}$}{#2}}}

\def\fpage#1{\begingroup
\voffset=.3in
\thispagestyle{empty}\begin{table}[b]\centerline{\footnotesize #1}
        \end{table}\endgroup}

\def\runninghead#1#2{\pagestyle{myheadings}
\markboth{{\protect\footnotesize\it{\quad #1}}\hfill}
{\hfill{\protect\footnotesize\it{#2\quad}}}}
\font\tenrm=cmr10
\font\tenit=cmti10
\font\tenbf=cmbx10
\font\bfit=cmbxti10 at 10pt
\font\ninerm=cmr9
\font\nineit=cmti9
\font\ninebf=cmbx9
\font\eightrm=cmr8

\def\qed{\hbox{${\vcenter{\vbox{                        
   \hrule height 0.4pt\hbox{\vrule width 0.4pt height 6pt
   \kern5pt\vrule width 0.4pt}\hrule height 0.4pt}}}$}}

\def\bsc{{\sc a\kern-6.4pt\sc a\kern-6.4pt\sc a}}       
\def\bflatex{\bf L\kern-.30em\raise.3ex\hbox{\bsc}\kern-.14em
T\kern-.1667em\lower.7ex\hbox{E}\kern-.125em X}

\newcommand{\be}{\begin{equation}}
\newcommand{\ee}{\end{equation}}
\newcommand{\bea}{\begin{eqnarray}}
\newcommand{\eea}{\end{eqnarray}}
\newcommand{\bd}{\begin{displaymath}}
\newcommand{\ed}{\end{displaymath}}
\newcommand{\prim}{{\scriptscriptstyle \prime}}
\newcommand{\x}{\mbox{x}}

\begin{document}

\runninghead{The use of Schoonschip and Form} {The use of Schoonschip
and Form}

\normalsize\textlineskip
\thispagestyle{empty}
\setcounter{page}{1}

\vspace*{-2.5cm}
{\flushright Preprint ROM2F/95/06}

{\flushright April 12, 1995}

\vspace*{0.59truein}

\fpage{1}
\centerline{\bf THE USE OF SCHOONSCHIP AND FORM}
\vspace*{0.035truein}
\centerline{\bf IN PERTURBATIVE LATTICE CALCULATIONS \footnote{Presented by
S.Capitani}}
\vspace*{0.37truein}
\centerline{\footnotesize STEFANO CAPITANI \footnote{Address from September
1995: Department of Physics, University of Wales, Swansea SA2 8PP, Wales,
United Kingdom}}
\vspace*{0.015truein}
\centerline{\footnotesize\it Dipartimento di Fisica, Universit\`a degli Studi
di Roma ``La Sapienza''}
\baselineskip=10pt
\centerline{\footnotesize\it P.le Aldo Moro 2, I-00185 Roma,
Italy}
\vspace*{0.03truein}
\centerline{\footnotesize\it INFN, Sezione di Roma 2}
\baselineskip=10pt
\centerline{\footnotesize\it Via della Ricerca Scientifica, I-00133 Roma,
Italy}
\vspace*{10pt}
\centerline{\normalsize and}
\vspace*{10pt}
\centerline{\footnotesize GIANCARLO ROSSI}
\vspace*{0.015truein}
\centerline{\footnotesize\it Dipartimento di Fisica, Universit\`a degli
Studi di Roma ``Tor Vergata''}
\baselineskip=10pt
\centerline{\footnotesize\it  INFN, Sezione di Roma 2}
\baselineskip=10pt
\centerline{\footnotesize\it Via della Ricerca Scientifica, I-00133 Roma,
Italy}
\vspace*{0.225truein}

\vspace*{0.21truein}
\abstracts{Using the formal languages Schoonschip and Form, we have
developed general codes that are able to carry out all the algebraic
manipulations needed to perform analytic lattice calculations, starting from
the elementary building blocks (propagators and vertices) of each Feynman
diagram. The main difficulty resides in the fact that, although there are
many built in instructions to deal with Dirac gamma-matrices, Schoonschip
and Form have been conceived having in mind a continuum theory, which is
invariant with respect to the Lorentz group. On the lattice, on the contrary,
a field theory is only invariant with respect to the hypercubic group,
contained in the (euclidean) Lorentz group and not every pair of equal
indices should be summed over. Being impossible to directly use the
`gammatrics' of Schoonschip and Form as they are, special routines
have been developed to correctly treat gamma matrices on the lattice,
while using as much as possible of the built in Schoonschip and Form
commands.}{We have used our codes to compute, in 1-loop perturbation theory
in lattice QCD, the renormalization constants and mixing coefficients
of the operators that enter in the determination of the first two moments
of deep inelastic scattering structure functions.}{}


\vspace*{1pt}\textlineskip      
\section{Introduction}    
\vspace*{-0.5pt}
\noindent
In this contribution we present a short report on the use of some algebraic
manipulation computer languages, namely Schoonschip and Form, to carry out
the large and complex manipulations which occur in certain analytic
perturbative lattice calculations. We have employed our codes to compute
in lattice QCD matrix elements of operators
related to the first two moments of the structure functions measured in
Deep Inelastic Scattering (DIS), both using the standard Wilson action and the
improved nearest-neighbor action. Up to now we have completed the calculations
of the 1-loop renormalization constants and mixing coefficients of the
fermion and gluon rank two operators ($O^q_{\mu \nu}$ and $O^g_{\mu \nu}$)
that are related to the first moment of DIS structure functions, and of the
rank
three operator ($O^q_{\mu \nu \tau}$) related to the second moment.

Consequently to the use of improvement in lattice QCD, there is indeed a
large number of diagrams to be computed and, in most of them, even in the
simplest case of the rank two operator, algebraic manipulations are very
complicated, each Feynman diagram giving rise to a huge number of terms.
For this reason, we have found unavoidable to check independently, by means
of suitable algebraic manipulation programs, all the computations made by hand.
For $O^q_{\mu \nu \tau}$ we almost completely have to rely on computer
results.

The first code we developed was written using the Schoonschip language, and was
used in the case of $O^q_{\mu \nu}$ and $O^g_{\mu \nu}$.\cite{capitani}
Afterwards, we upgraded this code to make it able to perform the more
complex calculations concerning $O^q_{\mu \nu \tau}$. In this case it is
an impossible task to calculate all diagrams by hand; the agreement for
the few of them that we could really compute by hand was nevertheless
complete.\cite{capitani2} Further checks were later offered by an independent
code developed by Beccarini using Form,\cite{beccarini0} and by another Form
code
constructed by us, that we have further upgraded to meet the requirements of
the higher complexity of the operators $O^q_{\mu \nu \tau}$ that we are
presently studying.\cite{beccarini}

\section{Moments of Structure Functions and Lattice QCD}
\noindent
In QCD the lowest twist operators appearing in the light-cone expansion
of the product of two hadronic weak currents, relevant to DIS, are
\bd
O^{qS}_{\mu_{1} \cdots \mu_{N}} = \frac{1}{2^N}
\overline{\psi} \: \gamma_{[ \mu_{1}} \!
\stackrel{\displaystyle \leftrightarrow}{D}_{\mu_{2}} \cdots
\stackrel{\displaystyle \leftrightarrow}{D}_{\mu_{N} ]} (1 \pm \gamma_5) \psi
\ed
\bd
O^{qNS}_{\mu_{1} \cdots \mu_{N}} = \frac{1}{2^N}
\overline{\psi} \: \gamma_{[ \mu_{1}} \!
\stackrel{\displaystyle \leftrightarrow}{D}_{\mu_{2}} \cdots
\stackrel{\displaystyle \leftrightarrow}{D}_{\mu_{N} ]} (1 \pm \gamma_5)
\frac{\lambda^f}{2} \psi
\ed
\be
O^{gS}_{\mu_{1} \cdots \mu_{N}} = \sum_{\rho} \mbox{Tr}
\left[ F_{[ \mu_{1} \rho}
\stackrel{\displaystyle \leftrightarrow}{D}_{\mu_{2}} \cdots
\stackrel{\displaystyle \leftrightarrow}{D}_{\mu_{N - 1}}
F_{\rho \mu_{N} ]} \right] \label{eq:opsns} ,
\ee
where the $\lambda^f$'s are flavor
matrices.\cite{part,gross,gross2,georgi,bardeen} The operators (\ref{eq:opsns})
are gauge invariant and all have twist two;\cite{twist} they are symmetrized
with respect to all Lorentz indices. $S$ and $NS$ superscripts refer to
Singlet and Non Singlet flavor structures.

The operators of rank two and three in (\ref{eq:opsns}) are related to the
first two moments of the DIS structure functions via the Wilson operator
expansion of the product of two electromagnetic currents (we do not consider
in this talk the $\gamma_5$ contributions).\cite{gross,Wilope} The knowledge
of the hadronic matrix elements of these operators is necessary for the
theoretical evaluation of the moments of the $\x$-distributions of quarks
and gluons inside the hadrons ($\x$ is the momentum fraction carried by
the struck parton inside the hadron).\cite{christ}

Our work consists in the determination of the 1-loop renormalization
constants and mixing coefficients of the lattice fermion and gluon operators
of rank two,\cite{capitani} and the renormalization constant
of the fermion operator of rank three that correspond to the continuum
expressions given in (\ref{eq:opsns}).\cite{beccarini} The calculations were
done using the nearest neighbor improved lattice QCD action and include the
results for the renormalization constants and mixing coefficient pertinent
to the standard Wilson action.\cite{qcd,Wil,Sym,Lus,She,rom1}
All these constants are needed to renormalize the lattice operators
and be able to extract physical hadronic matrix elements from numbers
obtained in Monte Carlo QCD simulations.

Lattice QCD represents today the only viable way of evaluating from first
principles the hadronic matrix elements needed for the computation of
the moments of the structure functions.
The nearest neighbor improved QCD action (also known as the ``clover-leaf''
action), is obtained by adding to the standard Wilson action (for one
flavor $f$ on a euclidean lattice)
\bd
S^{f}_{LATT} = a^4 \sum_{n\ } \Bigg[  - \frac{1}{2 a} \sum_{\mu} \Big[
\overline{\psi}_{n} ( r - \gamma_{\mu} ) U_{n,\mu} \psi_{n + \mu}
+ \overline{\psi}_{n + \mu} ( r + \gamma_{\mu} ) U_{n,\mu}^{+} \psi_{n}
\Big]
\ed
\be
+ \overline{\psi}_{n} \left( m_{f} + \frac{4 r}{a} \right) \psi_{n} \Bigg]
- {\displaystyle\frac{1}{g_0^2}} \sum_{n,\mu \nu}
\Bigg[ \mbox{Tr} \left[ U_{n,\mu} U_{n + \mu , \nu} U_{n + \nu , \mu}^{+}
U_{n, \nu}^{+} \right] - N_c \Bigg] \label{eq:wil},
\ee
the nearest-neighbor interaction term
\be
\Delta S^{f}_{I} = - i g_0 a^{4} \sum_{n,\mu \nu} \frac{r}{4 a} \:
\overline{\psi}_{n}
\sigma_{\mu \nu} F_{n, \mu \nu} \psi_{n} \label{eq:impr}.
\ee
Here $F_{n, \mu \nu}$ is not the naive lattice ``plaquette''
\be
P_{n, \mu \nu} = \frac{1}{2ig_0a^2} (U_{n, \mu \nu} - U^{+}_{n, \mu \nu})
\label{eq:effeno} ,
\ee
with $U_{n, \mu \nu} = U_{n,\mu} U_{n + \mu , \nu} U_{n + \nu , \mu}^{+}
U_{n, \nu}^{+}$,
but rather the average of the four plaquettes lying in the plane $\mu \nu$,
stemming from the point $n$:
\be
F_{n, \mu \nu} = \frac{1}{4} \sum_{\mu \nu = \pm} P_{n, \mu \nu} =
\frac{1}{8ig_0a^2} \sum_{\mu \nu = \pm} (U_{n, \mu \nu} - U^{+}_{n, \mu \nu}).
\label{eq:effesi}
\ee

The use of this action has been proved to remove from on-shell hadronic
matrix elements all terms that in the continuum limit are effectively of
order $a$,\cite{She,rom1,rom2,rom3} $a$ being the lattice spacing, provided
that
in the calculation of a fermionic Green function, each fermion field
undergoes the rotation
\be
\psi \longrightarrow \left( 1 - \frac{a r}{2} \stackrel{\displaystyle
\rightarrow }{\not\!\!{D}} \right) \psi \mbox{\ \ \ ,\ \ \ }
\overline{\psi} \longrightarrow \overline{\psi} \left( 1 + \frac{a r}{2}
\stackrel{\displaystyle \leftarrow }{\not\!\!{D}} \right) \label{eq:rotation}.
\ee
With these requirements, the difference between continuum and lattice
is lowered from
\be
\left< p \left| \widehat{\cal O}_{L} \right| p^{\prim} \right>_{Monte\ Carlo}=
a^{d} \left[ \left< p \left| \widehat{\cal O} \right| p^{\prim} \right>_{phys.}
+ O(a) \right]
\ee
(where $\left< p \left| \widehat{\cal O} \right| p^{\prim} \right>_{phys.}$
is the physical matrix element we want to extract from Monte Carlo data
and $d$ is its physical dimension) to
\be
\left< p \left| \widehat{\cal O}_{L} \right| p^{\prim}
\right>^{IMPR.}_{Monte\ Carlo}=
a^{d} \left[ \left< p \left| \widehat{\cal O} \right| p^{\prim} \right>_{phys.}
+ O(a/\log a) \right].
\ee
Using this recipe, the systematic error related to the lattice discretization
drops in many cases from 20 -- 30 percent down to 5 -- 10
percent.\cite{err1,err2}

The numbers we want to compute are the renormalization constants that
connect the bare lattice operators on the lattice, $O(a)$, to finite
operators, $\widehat{O}(\mu)$, renormalized at a scale $\mu$:
\be
\widehat{O}^l(\mu) = Z_{lk}(\mu a) O^k(a) \label{eq:cosrin}.
\ee
The constants $Z_{lk}$ are fixed in perturbation theory by the same
renormalization conditions used in the continuum.
In the flavor Singlet case there is a mixing between quark and gluon operators
of rank two that have the same conserved quantum numbers.
However, with the choice $\mu \neq \nu$ and the
definition (\ref{eq:effesi}) of the gauge field strength we avoid mixing with
operators of lower dimensions, and therefore the need for
subtractions of power divergences. We thus write:
\bd
\widehat{O}^q = Z_{qq} O^q + Z_{qg} O^g
\ed
\be
\widehat{O}^g = Z_{gq} O^q + Z_{gg} O^g \label{eq:ref1},
\ee
where the $Z$'s are determined by imposing the renormalization conditions:
\bea
<q| \widehat{O}^q (\mu) |q> & = & <q| O^q (a) |q> |^{tree}_{p^2 = \mu^2}
\nonumber \\
<g,\sigma| \widehat{O}^q (\mu) |g,\sigma>
& = & <g,\sigma| O^q (a) |g,\sigma> |^{tree}_{p^2 = \mu^2} = 0  \nonumber \\
<q| \widehat{O}^g (\mu) |q>
& = &  <q| O^g (a) |q> |^{tree}_{p^2 = \mu^2} = 0 \nonumber \\
<g,\sigma| \widehat{O}^g (\mu) |g,\sigma>
& = & <g,\sigma| O^g (a) |g,\sigma> |^{tree}_{p^2 = \mu^2} \label{eq:ref2},
\eea
where $|q>$ and $|g,\sigma>$ are respectively quark and gluon states.

Similarly in the case of $O^q_{\mu \nu \tau}$ there is no mixing with lower
dimensional operators, if one chooses $\mu \neq \nu \neq \tau$. In fact in
the ``quenched'' approximation (no internal quark loops) $O^q_{\mu \nu}$
and $O^q_{\mu \nu \tau}$ are simply multiplicatively renormalizable (no
mixing with other operators).

\section{What our Programs do}
\noindent
We have developed general codes able to automatically carry out all the
necessary algebraic manipulations, starting from the elementary building
blocks of the calculation represented by the expressions of propagators,
vertices and Fourier transforms of the operators. The final output of our
codes can be cast either in the form of an analytic expression (to be
possibly compared with the calculations made by hand) or in a format
ready to be introduced in a suitable Fortran code for the final numerical
loop integration.

The main difficulty in using Schoonschip and Form in lattice calculations
resides in the fact that, although these languages
have many built in instructions to deal with gamma matrices, they have been
conceived having in mind a continuum theory, which is invariant with respect
to the (euclidean) Lorentz group $O(4)$. On the lattice, on the contrary, where
the theory is only invariant with respect to the hypercubic group
$H(4)$,\cite{mix1,mix2} we cannot use the usual summation conventions and
the standard rules to rearrange gamma matrices. To make an example, one
simple and frequently occurring term, like
$\sum_{\lambda} \gamma_{\lambda} p_{\lambda} \sin k_{\lambda}$, is not
properly handled by these languages. Actually, this term is wrongly
reduced by Schoonschip and Form to $\not\!\!{p} \sin k_{\lambda}$, because
the first two equal indices that are encountered are by default assumed to be
contracted. It turns out that most of the terms that arise on the lattice are
improperly handled by a straight use of the Schoonschip and Form commands.

Being impossible to directly use the ``gammatrics'' of Schoonschip
and Form as they are, it has been necessary to develop special routines
to correctly treat gamma matrices on the lattice, while using as much
as possible of the built in Schoonschip and Form commands. From our efforts
at least three sets of routines have grown up, and we have checked for each
of them that the rather complicated Dirac algebra (we have products of up to
seven gamma matrices) is carried out correctly.

The CPU times needed to run our programs has greatly shrunk with the
evolution of the various codes developed in the last three years.
The first Schoonschip code we used in the case of the rank two operator
was running on a Sun 3 workstation, with CPU times that varied, depending
on the complexity of the diagrams, from 20 seconds up to 5 minutes for
the most complicated cases.\cite{capitani} This was quite reasonable
for our purposes, but when turning to the rank three operator the time
required by the appropriately modified Schoonschip code was one order
of magnitude higher.\cite{capitani2}
More recently we have developed new codes written using Form that run on
VAX-VMS machines. These programs are considerably faster, and the
typical CPU times are back in the minute range. There has been a final
reduction of these values when we have turned to an HP-UX 9000/735
machine.\cite{beccarini}

Another big problem we have encountered was the limitation on the working
memory allowed by Schoonschip and Form on the different machines we have used.
This is of primary relevance in the first stages of the computation, when all
vertices, propagators etc. are expanded, up to the first and second order in
the lattice spacing $a$, for the rank two and three operators respectively.
In our case, when dealing with the rank three case, we have several products
of up to ten trigonometric functions, each one to be expanded to second order
in $a$. Their product thus contains tenths of thousands of monomials. A large
part of them does not contribute to the final expression, to the order in $a$
we are interested in, and has to be killed at the earliest possible stage
of the calculation.

As a last remark, we want to say that, besides being more portable on
different machines, Form appears to be superior to Schoonschip,
at least for our kind of calculations (not involving $\gamma_5$ and limited
to four dimensions). Form seems indeed to be faster and to allow an easier
localization of the errors. We have also found very useful the larger set of
wildcards that Form owns compared to Schoonschip.

\nonumsection{Acknowledgements}
\noindent
S.C. acknowledges the EC-HMC contract no. CHRX-CT92-0051 for supporting
his participation to the IV International Workshop AIHENP.

\nonumsection{References}


\noindent

\begin{thebibliography}{000}
\bibitem{capitani}
S.Capitani and G.C.Rossi, {\bibit Nucl. Phys.} {\bibbf B433}, 351 (1995)
\bibitem{capitani2}
S.Capitani, Ph.D. Thesis, Univ. of Rome ``La Sapienza'', Rome, 1994
\bibitem{beccarini0}
G.Beccarini, Undergraduate Thesis, Univ. of L'Aquila, L'Aquila, 1993
(1993)
\bibitem{beccarini}
G.Beccarini, S.Capitani and G.C.Rossi, in preparation
\bibitem{part}
G.Altarelli, {\bibit Phys. Rep.} {\bibbf 81}, 1 (1982)
\bibitem{gross}
D.J.Gross, in {\bibit Methods in field theory --- Les Houches, Session
XXVIII, 1975}, eds. R.Balian and J.Zinn-Justin (North-Holland,
Amsterdam, 1976), p. 141
\bibitem{gross2}
D.J.Gross and F.Wilczek, {\bibit Phys. Rev.} {\bibbf D8}, 3633 (1973)
and {\bibit Phys. Rev.} {\bibbf D9}, 980 (1974)
\bibitem{georgi}
H.Georgi and H.D.Politzer, {\bibit Phys. Rev.} {\bibbf D9}, 416 (1974)
\bibitem{bardeen}
W.A.Bardeen, A.J.Buras, D.W.Duke and T.Muta, {\bibit Phys. Rev.} {\bibbf D18},
3998 (1978)
\bibitem{twist}
D.J.Gross and S.B.Treiman, {\bibit Phys. Rev.} {\bibbf D4}, 1059 (1971)
\bibitem{Wilope}
K.G.Wilson, {\bibit Phys. Rev.} {\bibbf 179}, 1499 (1969)
\bibitem{christ}
N.Christ, B.Hasslacher and A.Mueller, {\bibit Phys. Rev.} {\bibbf D6},
3543 (1972)
\bibitem{qcd}
H.Fritzsch, M.Gell-Mann and H.Leutwyler, {\bibit Phys. Lett.} {\bibbf B47},
365 (1973)
\bibitem{Wil}
K.G.Wilson, {\bibit Phys. Rev.} {\bibbf D10}, 2445 (1974), and in
{\bibit New Phenomena in Subnuclear Physics}, ed. A.Zichichi
(Plenum Press, New York, 1977)
\bibitem{Sym}
K.Symanzik, in {\bibit Mathematical Problems in Theoretical
Physics}, ed. R.Schrader {\bibit et al.}, Springer Lecture Notes in Physics,
vol. {\bibbf 153}, p. 47 (1982)
\bibitem{Lus}
M.L\"uscher and P.Weisz, {\bibit Commun. Math. Phys.} {\bibbf 97}, 59 (1985)
\bibitem{She}
B.Sheikholeslami and R.Wohlert, {\bibit Nucl. Phys.} {\bibbf B259}, 572 (1985)
\bibitem{rom1}
G.Heatlie, G.Martinelli, C.Pittori, G.C.Rossi and C.T.Sachrajda,
{\bibit Nucl. Phys.} {\bibbf B352}, 266 (1991)  and
{\bibit Nucl. Phys.} {\bibbf B} {\bibit (Proc. Suppl.)} {\bibbf 17}, 607
(1990);
E.Gabrielli, G.Heatlie, G.Martinelli, C.Pittori, G.C.Rossi, C.T.Sachrajda and
A.Vladikas, {\bibit Nucl. Phys.} {\bibbf B} {\bibit (Proc. Suppl.)} {\bibbf
20},
448 (1991)
\bibitem{rom2}
E.Gabrielli, G.Martinelli, C.Pittori, G.Heatlie and
C.T.Sachrajda, Nucl. Phys. {\bf B362} (1991) 475
\bibitem{rom3}
R.Frezzotti, E.Gabrielli, C.Pittori and G.C.Rossi, Nucl. Phys.
{\bf B373} (1992) 781
\bibitem{err1}
G.Martinelli, C.T.Sachrajda, G.Salina and A.Vladikas,
{\bibit Nucl. Phys.} {\bibbf B378}, 592 (1992); E: {\bibit Nucl. Phys.}
{\bibbf B397}, 479 (1993)
\bibitem{err2}
G.Martinelli, S.Petrarca, C.T.Sachrajda and A.Vladikas,
{\bibit Phys. Lett.} {\bibbf B311}, 241 (1993)
\bibitem{mix1}
M.Baake, B.Gem\"unden and R.Oedingen, {\bibit Journ. Math. Phys.}
{\bibbf 23}, 944 (1982)
\bibitem{mix2}
J.Mandula, G.Zweig and J.Govaerts, {\bibit Nucl. Phys.} {\bibbf B228}, 91
(1983)
\end{thebibliography}
\end{document}